\documentclass[prc,a4paper,twocolumn,amsmath,amssymb,showpacs,showkeys,floatfix]{revtex4}
\usepackage[latin1]{inputenc}
\usepackage[T1]{fontenc}
\usepackage{units}
\usepackage{graphicx}
\usepackage{dcolumn}

\begin{document}

\title{Improved short-range correlations and $0\nu\beta\beta$
  nuclear matrix elements of $^{76}$Ge and $^{82}$Se}
\author{Markus Kortelainen}
\author{Jouni Suhonen}
\affiliation{Department of Physics, University of Jyv\"{a}skyl\"{a}, 
P.O.\ Box 35, FIN-40351 Jyv\"{a}skyl\"{a}, Finland}

\begin{abstract}
We calculate the nuclear matrix elements of the neutrinoless double
beta ($0\nu\beta\beta$) decays of $^{76}$Ge and $^{82}$Se for the
light-neutrino exchange mechanism. The nuclear wave functions are obtained
by using realistic two-body forces within the proton-neutron quasiparticle
random-phase approximation (pnQRPA). We include the effects that come from
the finite size of a nucleon, from the higher-order terms of nucleonic 
weak currents, and from the nucleon-nucleon short-range correlations. 
Most importantly, we improve on the presently available calculations 
by replacing the rudimentary Jastrow short-range correlations by the 
more advanced unitary correlation operator method (UCOM). The UCOM 
corrected matrix elements turn out to be notably larger in magnitude 
than the Jastrow corrected ones. This has drastic consequences for 
the detectability of $0\nu\beta\beta$ decay in the present and future 
double beta experiments.
\end{abstract}


\maketitle

The nuclear matrix elements of neutrinoless double beta
($0\nu\beta\beta$) decay have become an important issue in the
present-day neutrino physics (see, e.g., \cite{ELL02, ELL04, BIL04, GEH07}). 
This has been boosted by the
verification of the existence of neutrino mass by the oscillation
experiments \cite{Oscillations} and the claimed discovery of the 
$0\nu\beta\beta$ decay \cite{KLA01,KLA04}. Further incentive to
produce reliable nuclear matrix elements comes from the needs of the
running NEMO 3 \cite{ARN05a} and CUORICINO \cite{ARN05b} experiments,
as well as the future large-scale experiments under R\&D planning and 
construction (see, e.g., \cite{BAR04}). For these experiments the 
nuclear matrix elements are an essential prerequisite in extraction 
of reliable values for the absolute mass scale of the neutrino
\cite{DOI85} and possibly the CP phases of the neutrino-mixing matrix
\cite{PAS02}. 

The most popular nuclear model to treat the structure of medium-heavy
and heavy double-beta decaying nuclei is the proton-neutron quasiparticle
random-phase approximation (pnQRPA) \cite{TOM91,SUH98,FAE98}. Also
some recent shell-model results are available \cite{CAU99,CAU05}.
The pnQRPA is a model tailored to describe, in an efficient way, the 
energy levels of odd-odd nuclei and their beta decays to the
neighboring even-even nuclei \cite{SUH07}. Also its derivative,
renormalized pnQRPA \cite{TOI95}, has been used \cite{TOI97,ROD06} to
compute double-beta matrix elements, although its use has been heavily
criticized (see, e.g., \cite{RQRPA} and references therein). A problem
with the use of the pnQRPA (and the renormalized pnQRPA) 
is that it contains a free parameter, the so-called 
particle-particle strength parameter, $g_{\rm pp}$, that
controls the magnitude of the proton-neutron two-body interaction 
matrix elements in the $T=0$ pairing channel \cite{VOG86,CIV87}. There
are basically two ways to fix the value of this parameter, either by using
the data on two-neutrino double beta ($2\nu\beta\beta$) decay
\cite{ROD06} or the data on single beta decay \cite{SUH05,CIV05}. In the
case of $^{76}$Ge and $^{82}$Se there is no available data
on single beta decays so that in this work we have chosen to use the
$2\nu\beta\beta$ data to fix the possible values of $g_{\rm pp}$.

In this article we address the mass mode of the $0\nu\beta\beta$ decay
where a light virtual Majorana neutrino is exchanged by the two decaying
neutrons of the initial nucleus. Typically the exchanged momentum is
so large as to force the two neutrons to overlap unless steps are
taken to prevent the occurance of such a spurious event. The
traditional way \cite{HAX84,ENG88} to remove this spuriosity is to introduce 
an explicit Jastrow type of correlation function into the involved two-body
transition matrix elements in the parametrization of Miller and
Spencer \cite{MIL76}. This method, although microscopically inspired,
is just a phenomenological way to introduce short-range correlations
into the two-nucleon relative wave function. A conspicuous flaw of
the Jastrow method is that the Jastrow function effectively cuts out the 
small $r$ \cite{r} part from the relative wave function of the two nucleons.
For this reason, the traditionally adopted Jastrow procedure \cite{TOM91}
does not conserve the norm of the relative wave function \cite{rudi}. 

In the present calculations we improve on the Jastrow method and adopt
the more sophisticated microscopic approach of unitary correlation operator 
method (UCOM) \cite{FEL98}. In the UCOM one obtains the 
correlated many-particle state $\vert \tilde{\Psi} \rangle$ from 
the uncorrelated one as
\begin{equation}
\vert \tilde{\Psi} \rangle = C \vert\Psi\rangle \, ,
\end{equation}
where $C$ is the unitary correlation operator \cite{FEL98}. Due to the 
unitarity of the operator $C$, the norm of the correlated state is 
conserved and no amplitude is lost in the relative wave function. 
In the $0\nu\beta\beta$ calculations this leads to a more complete
description of the relative wave function at small distances
$r$, as was demonstrated in \cite{KOR07}. It should be stressed that
no extra free parameters are introduced by the use of the UCOM
at the level of double-beta-decay calculations. All the needed
parameters have been fixed by minimization procedures for
$\beta\beta$-independent observables \cite{ROT05}.
The UCOM method has been demonstrated \cite{ROT06} to produce good results
for the binding energies of nuclei over a wide mass range already 
at the Hartree-Fock level. It was also shown that the UCOM renders a
good starting point for inclusion of long-range correlations by means
of many-body perturbation theory.
 
In \cite{KOR07} it was demonstrated for the $^{48}$Ca and $^{76}$Ge 
$0\nu\beta\beta$ decays that the Jastrow procedure leads to the 
excessive reduction of 30\% -- 40\% in the magnitudes of the 
$0\nu\beta\beta$ nuclear matrix elements. At the same time the UCOM 
reduces the magnitudes of the matrix elements only by 7\% -- 16\%. 
This explains the large short-range correlation corrections to the
matrix elements of \cite{ROD06}. 

The issue about the magnitude of the short-range corrections is an
extremely important one since it directly affects the magnitudes of
the relevant nuclear matrix elements used to extract the neutrino
masses from potentially succesful future double beta experiments.
There are large differences between the Jastrow and UCOM corrections,
the Jastrow corrected matrix elements being substantially smaller
that the UCOM corrected ones. This difference would severely alter the
predicted sensitivities of future neutrino experiments, the UCOM
corrected matrix elements being more favourable for the detection of
neutrinoless double beta decay.

The double beta decays of $^{76}$Ge and $^{82}$Se proceed through the
virtual states of the intermediate nuclei $^{76}$As and $^{82}$Br to
the ground states of the final nuclei $^{76}$Se and $^{82}$Kr.
By assuming the neutrino-mass mechanism to be the dominant one, we can 
write the inverse of the half-life as \cite{SUH98}
\begin{equation} \label{eq:half}
\left[ t_{1/2}^{(0\nu)}\right]^{-1} = G_{1}^{(0\nu)}
\left( \frac{\langle m_{\nu}\rangle }{m_{\rm e}} \right)^{2}
\left( M^{(0\nu)}\right)^{2}\, ,
\end{equation}
where $m_{\rm e}$ is the electron mass and $G_{1}^{(0\nu)}$
is the leptonic phase-space factor. The $0\nu\beta\beta$
nuclear matrix element $M^{(0\nu)}$ consists of the
Gamow--Teller, Fermi and tensor parts as
\begin{equation} \label{eq:me}
M^{(0\nu)} = M_{\rm GT}^{(0\nu)} - \left( \frac{g_{\rm V}}{g_{\rm A}}
\right)^{2} M_{\rm F}^{(0\nu)} + M_{\rm T}^{(0\nu)} \ .
\end{equation}

Numerical calculations show that the tensor part in (\ref{eq:me}) 
is quite small and its contribution can be safely neglected in what follows.
The expressions for the phase-space factor, the effective neutrino 
mass $\langle m_{\nu}\rangle$ and the matrix elements of (\ref{eq:me}) 
are given, e.g., in \cite{DOI85,TOM91,SUH98}. To the ``bare''
matrix elements we have applied the Jastrow short-range
correlation corrections, together with the 
\emph{higher-order terms of nucleonic weak currents} and the
\emph{nucleon's finite-size} corrections in the way described in 
\cite{ROD06,SIM99}. In addition, we have computed the corrected 
matrix elements by replacing the Jastrow correlations by the UCOM 
correlations.

We calculated the wave functions of all the nuclear states in the 
intermediate nuclei by the use of the pnQRPA framework in the model 
space of 1p-0f-2s-1d-0g-0h$_{11/2}$ single-particle orbitals, both for
protons and neutrons. The single-particle energies were obtained from
a spherical Coulomb-corrected Woods--Saxon potential with a standard
parametrization, optimized for nuclei near the line of beta
stability. Slight adjustments were done for some of the
energies at the vicinity of the proton and neutron Fermi
surfaces to reproduce better the low-energy spectra of the neighboring
odd-$A$ nuclei and those of the intermediate nuclei.

\begin{figure}[htb]
\includegraphics[width=9cm]{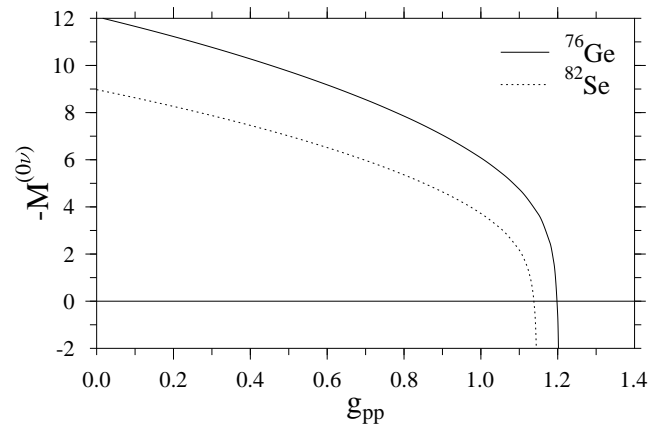}
\caption{The $0\nu\beta\beta$ nuclear matrix elements $M^{(0\nu)}$ 
  of Eq.~(\ref{eq:me}) for the decays of $^{76}$Ge and $^{82}$Se as
  functions of the particle-particle interaction parameter $g_{\rm pp}$.}
\label{fig:gppme}
\end{figure}

The Bonn-A G-matrix was used as a two-body interaction and it was 
renormalized in the standard way, as discussed e.g. in Refs. 
\cite{SUH88,SUH93}. Due to this phenomenological renormalization we did 
not perform an additional UCOM renormalization \cite{ROT05} of the 
two-body interaction. After fixing all the Hamiltonian parameters the
only free parameter left was the $g_{\rm pp}$ parameter mentioned
earlier. In Fig.~\ref{fig:gppme} we have studied the $g_{\rm pp}$ 
dependence of the matrix element $M^{(0\nu)}$ of
Eq.~(\ref{eq:me}) for both $^{76}$Ge and $^{82}$Se. We have used
$R=1.2A^{1/3}\,\textrm{fm}$ as the nuclear radius, and the finite-size, 
higher-order-term and UCOM corrections were taken into account.
Here one can see the typical break-down of the pnQRPA at
large values of $g_{\rm pp}$. Moreover, the corresponding
break-down points for the two nuclei are close to each other.

We obtained the physical values of $g_{\rm pp}$ by using the
method of \cite{ROD06}. Consequently, we used the recommended data
\cite{BAR06} on $2\nu\beta\beta$-decay half-lives of $^{76}$Ge and
$^{82}$Se by including the experimental error limits and the
uncertainty in the value of the axial-vector coupling constant 
$1.0\le g_{\rm A}\le 1.254$. The resulting intervals \cite{posint} of 
``experimental matrix elements'' were then converted to the following
intervals of $g_{\rm pp}$ values
\begin{equation} 
\begin{split}
\label{eq:intervals}
&1.02\le g_{\rm pp}\le 1.06 \ \textrm{ for } \,^{76}\textrm{Ge} \ ,\\
&0.96\le g_{\rm pp}\le 1.00 \ \textrm{ for } \,^{82}\textrm{Se} \ .
\end{split}
\end{equation}

\begin{figure}[htb]
\includegraphics[width=9cm]{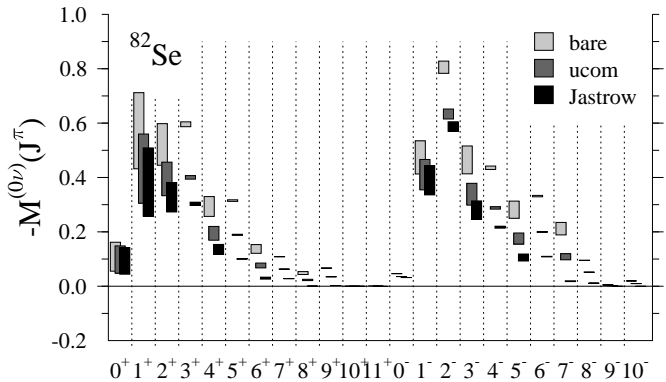}
\caption{Multipole decomposition of the 
  matrix element $M^{(0\nu)}$ of Eq.~(\ref{eq:me}) for the decay 
  of $^{82}$Se. The spread in the multipole contributions corresponds
  to the $g_{\rm pp}$ interval for $^{82}$Se in (\ref{eq:intervals}).}
\label{fig:decom}
\end{figure}

In Fig.~\ref{fig:decom} we show for the $^{82}$Se decay the
decomposition of the total matrix element (\ref{eq:me}) into
multipoles. The bare matrix element contains no short-range, 
finite-size or higher-order-term corrections, whereas the UCOM and 
Jastrow matrix elements include all 
these corrections. The spread in the multipole contributions corresponds
to the $g_{\rm pp}$ interval for $^{82}$Se in (\ref{eq:intervals}). 
More specifically, the upper end of the bar represents the case 
$g_{\rm pp}=0.96$) with $g_{\rm A}=1.0$ and the lower end the case 
$g_{\rm pp}=1.00$ with $g_{\rm A}=1.254$. In
the figure we see that the corrections substantially reduce the
magnitude of a given multipole contribution. The differences between
the Jastrow and UCOM corrected multipole contributions increase with 
increasing multipole. This pattern is reminiscent of the
one shown in Fig.~3 of \cite{KOR07} where no finite-size or 
higher-order-term corrections were taken into account. All this bears
evidence of the fact that the rudimentary Jastrow method overestimates the 
effect of nuclear short-range correlations in $0\nu\beta\beta$-decay 
calculations. 

\begin{table}
\caption{\label{tab:corr} Matrix element $M^{(0\nu)}$ of
  Eq.~(\ref{eq:me}) computed by correcting successively the bare
  matrix element (b.m.e.) by the higher-order terms in the nucleonic 
  current (A), by the nucleon finite-size effect (B), and by either
  the Jastrow (C) or UCOM (D) correlations. The values $g_{\rm pp}=1.0$
  and $g_{\rm A}=1.254$ were used in the calculations.}
\begin{ruledtabular}
	\begin{tabular}{lccccc}
Nucleus	& b.m.e. & +A & +A+B & +A+B+C & +A+B+D \\
	\hline
$^{76}$Ge & $-8.529$ & $-7.720$ & $-6.356$ & $-4.723$ & $-6.080$ \\
$^{82}$Se & $-5.398$ & $-4.862$ & $-3.914$ & $-2.771$ & $-3.722$ \\
	\end{tabular}
\end{ruledtabular}
\end{table}

In Table~\ref{tab:corr} we display the effects of the various
corrections to the matrix elements $M^{(0\nu)}$ of $^{76}$Ge and $^{82}$Se.
There we show in the second column the bare matrix elements (b.m.e.), 
then in the third column we show the b.m.e. corrected for the
higher-order terms in the nucleonic current (A). In the fourth column
we have added the nucleon finite-size effect (B) to the previous
matrix elements (b.m.e.+A), and finally, in the last two columns, we
have added to the previous matrix elements (b.m.e.+A+B) either the
Jastrow (C) or UCOM (D) short-range corrections.

\begin{table}
\caption{\label{tab:comp1} Matrix element $\vert M^{(0\nu)}\vert$ of
  Eq.~(\ref{eq:me}) for $^{76}$Ge obtained in the present calculation 
  and by \v Simkovic {\it et al.} \cite{SIM99}. Shown are the results 
  without and with the short-range correlations (s.r.c.) for
  $g_{\rm A}=1.254$.}
\begin{ruledtabular}
	\begin{tabular}{ccccc}
\multicolumn{2}{c}{without s.r.c.} & \multicolumn{3}{c}{with s.r.c.}	\\
Present & \cite{SIM99} & Jastrow & UCOM & \cite{SIM99} \\
	\hline
6.36 & 5.16 & 4.72 & 6.08 & 2.80 \\
	\end{tabular}
\end{ruledtabular}
\end{table}

Table~\ref{tab:corr} is very interesting in the sense that there we
can access the magnitudes of the various corrections to the bare
matrix element. The magnitudes of the corrections coming from the 
finite nucleon size and the higher-order terms of the nucleonic 
current together amount to 25\% -- 30\%. 
In fact, the magnitude of our bare matrix element is
rougly the one reported in \cite{ROD06,SIM99}. Even after correcting
by the higher-order terms and the nucleon finite-size effect the
matrix elements agree roughly, as shown in Table~\ref{tab:comp1}. 
From the table we also see that our computed Jastrow
corrections are much less than the ones obtained in 
\cite{ROD06,SIM99}. The reason for this is not clear. The net effect is that 
our Jastrow, and especially the UCOM
corrected matrix elements are much larger than the ones of 
\cite{ROD06,SIM99}. On the other hand, our shell-model computed Jastrow 
correlation corrections for $^{48}$Ca \cite{KOR07} agree with the ones
of \cite{RET95}.

\begin{table}
\caption{\label{tab:beta} Beta decay $\log ft$ values for transitions
  from the $2^-_1$ states of $^{76}$As and $^{82}$Br to one- and
  two-phonon states in the indicated final nuclei.}
\begin{ruledtabular}
	\begin{tabular}{lcccc}
& \multicolumn{2}{c}{$^{76}$Se}	& \multicolumn{2}{c}{$^{82}$Kr}	\\
$J_f$	& exp. & th. & exp. & th. \\
	\hline
$0^+_{\rm g.s.}$	& 9.7 &	9.0 & 8.9 & 9.3 \\
$2^+_1$ & 8.1 &	7.7 & 7.9 & 7.7 \\
$0^+_2$ & 10.3 & 9.2 & $\ge 9.6$ & 9.4 \\
$2^+_2$ & 8.2 & 8.7 & 8.0 & 9.0 \\
$4^+_1$ & 11.1 & 10.9 &	? & 11.1 \\
	\end{tabular}
\end{ruledtabular}
\end{table}

To check the consistency of our calculations we also computed the
single $\beta^-$ decay rates from the lowest $2^-$ states of the
intermediate nuclei $^{76}$As and $^{82}$Br to the lowest $2^+$ 
collective state, $2^+_1$, and its $0^+_2$, $2^+_2$ and $4^+_1$ 
two-phonon excitations (see, e.g., \cite{SUH93}) in 
$^{76}$Se and $^{82}$Kr. The wave function of the  
$2^+_1$ state was calculated by the use of the quasiparticle
random-phase approximation (QRPA) \cite{SUH07} and its energy was
fixed to the experimental one \cite{SUH93}. Beta decays to the
mentioned final states were computed by the method of the multiple
commutator model (MCM) of \cite{SUH93}. 

The $2^-$ wave function was calculated by using the central value 
of $g_{\rm pp}$ in the intervals (\ref{eq:intervals}). In the case of
a $2^-$ initial state this choice works fine since the calculated
beta-decay rates depend only weakly on $g_{\rm pp}$ within the 
range relevant for the $2\nu\beta\beta$ and $0\nu\beta\beta$ decays. 
We compare the computed $\log ft$ values with the available data in 
Table~\ref{tab:beta}. From this table it is seen that the computed 
numbers nicely reproduce the trends of the measured ones, although 
the assumpion that the states $0^+_2$, $2^+_2$ and $4^+_1$ are pure
two-phonon excitations is an idealized one. It has to be noted that
there is no experimental data on beta decay of the lowest $1^+$ state
in $^{76}$As and $^{82}$Br. For some double-beta-decaying systems the $1^+$
data exists and matching of beta and double beta decay could be more
problematic due to the stronger $g_{\rm pp}$ dependence of the
beta-decay rates from a $1^+$ intermediate state \cite{SUH05}.

\begin{table}
\caption{\label{tab:finalme} Nuclear $0\nu\beta\beta$ matrix elements 
  of Eq.~(\ref{eq:me}) for the decays of $^{76}$Ge and
  $^{82}$Se. The UCOM and other corrections are included. The used 
  $g_{\rm pp}$ values are also indicated.}
\begin{ruledtabular}
	\begin{tabular}{lcccc}
& \multicolumn{2}{c}{$^{76}$Ge}	& \multicolumn{2}{c}{$^{82}$Se}	\\
$g_{\rm pp}$ & 1.02 & 1.06 & 0.96 & 1.0 \\
\hline
$M^{(0\nu )}_{\rm F}$  &  1.923 &  1.803 &  1.304 &  1.214 \\
$M^{(0\nu )}_{\rm GT}$ & $-4.632$ & $-4.208$ & $-3.293$ & $-2.950$ \\
$M^{(0\nu )}$ & $-6.555$ & $-5.355$ & $-4.597$ & $-3.722$ \\
	\end{tabular}
\end{ruledtabular}
\end{table}

Our final results for the $0\nu\beta\beta$ nuclear matrix elements
have been collected in Table~\ref{tab:finalme}. These
matrix elements were calculated with the UCOM short-range corrections
by taking into account also the finite-size of the nucleons and the higher-order 
terms in the nucleonic weak current. The two different values of the
matrix elements correspond to the $g_{\rm pp}$ and $g_{\rm A}$
parameter combinations indicated earlier in the text. As can be seen,
the values of the final matrix elements vary between 
\begin{equation}
\begin{split} 
\label{eq:finalme}
&5.355\le\vert M^{(0\nu )}\vert\le 6.555 \ \textrm{ for }
\,^{76}\textrm{Ge} \ , \\
&3.722\le\vert M^{(0\nu )}\vert\le 4.597 \ \textrm{ for }
\,^{82}\textrm{Se} \ .
\end{split}
\end{equation}

We compare these matrix elements with other recent calculations in
Table~\ref{tab:comp2}. There the values $1.0\le g_{\rm A}\le 1.254$
are used for the axial-vector coupling constant, except that
\v Simkovic {\it et al.} \cite{SIM99} use $g_{\rm A}=1.254$. The
results of Civitarese {\it et al.} \cite{CIV05} are based on the
formalism introduced in \cite{SUH90} where the finite-size of the 
nucleon and the nucleonic weak current were obtained from a
relativistic quark-confinement model. In \cite{SUH90} the generated
nucleonic weak current is incomplete as compared to the present
formalism, adopted from \cite{SIM99}. Also the short-range correlations
were not taken into account. In this sense the last column of
Table~\ref{tab:comp2} should be compared to the fourth column ``+A+B''
of Table~\ref{tab:corr}. As already said, the differences between the
two results can be explained by the different treatment of the
weak-interaction current and the nucleon form factor.
The above stated, we conclude that our present results are more
complete that the ones of \cite{CIV05} and thus should be more reliable.
It is worth mentioning that in \cite{KOR07} the quoted matrix elements
were calculated for the ``default'' value $g_{\rm pp}=1.00$ without
taking into account the higher-order terms in the nucleonic current 
and the nucleon finite-size effect.

\begin{table}
\caption{\label{tab:comp2} Values of the matrix element 
  $\vert M^{(0\nu)}\vert$ of Eq.~(\ref{eq:me}) obtained in several 
  recent calculations.}
\begin{ruledtabular}
	\begin{tabular}{ccccc}
Nucleus	& Present & \cite{ROD06} & \cite{SIM99} & \cite{CIV05} \\
	\hline
$^{76}$Ge & $5.36-6.56$ & $2.26-2.74$ & 2.80 & $4.03-5.92$ \\
$^{82}$Se & $3.72-4.60$ & $1.86-2.45$ & 2.64 & $2.82-4.14$ \\
	\end{tabular}
\end{ruledtabular}
\end{table}

In the matrix element calculations there may be other 
uncertainties than the ones induced by the uncertainty in the value 
of $g_{\rm pp}$. Such uncertainties could arise from sources such as
deformation, the mean-field single-particle energies, and the adopted
two-body interaction. In \cite{ROD06} it was shown that the effect of
the adopted two-body interaction is very small as long as the
interaction is microscopic. Our adopted Bonn-A interaction is of this
type and included in the survey of \cite{ROD06}. In \cite{ROD06}
it was furthermore demonstrated that the size of the single-particle
space does not produce sizable effects as long as the value of
$g_{\rm pp}$ is determined from the $2\nu\beta\beta$ data, as is done
in the present calculations. By the same argument, only small effects
are expected from different parametrizations of the Woods-Saxon
mean-field potential and the resulting slightly different 
single-particle energies. These two sources of uncertainty produce 
effects that can be expected to be smaller than the one coming from 
the short-range correlations. 

The role of deformation is the most uncertain one. The presently discussed
nuclei of the two double-beta-decay chains are pf-shell nuclei and
most likely they have no or very small static deformation. Instead,
they are most likely soft anharmonic vibrators. The deformation allows of
a new suppression mechanism of $2\nu\beta\beta$ decay, namely through the
overlap factor used to take into account the non-orthogonality of the
intermediate states generated by using the initial and final ground
states as starting points in two separate pnQRPA calculations. 
This suppression mechanism is
enhanced when the deformations of the initial and final nuclei of 
double beta decay are different \cite{ALV04}. However, in \cite{FRE07}
it was deduced experimentally that $^{76}$Ge and $^{76}$Se exhibit
quantitatively very similar neutron pair correlations. This would
indicate similarity of their ground states and no suppression would arise
through different ground-state deformations. For $^{82}$Se and
$^{82}$Kr this question is still open. In any case, the role of deformation in 
$0\nu\beta\beta$ decay is still largely unexplored and no definitive
conclusion about the importance of deformation can be drawn for the present.

Our final matrix elements can be converted to $0\nu\beta\beta$
half-lives by choosing a value for the effective neutrino mass in
(\ref{eq:half}). Expressing the effective mass in units of eV and
using the phase-space integrals tabulated in \cite{SUH98} we obtain
for the predicted half-lives
\begin{equation} 
\begin{split} 
\label{eq:halflives}
&t_{1/2}^{(0\nu)} = (0.96-1.44)\times 10^{24}\,\textrm{yr}/
(\langle m_{\nu}\rangle[\textrm{eV}])^2 \textrm{ for}\ \,^{76}\textrm{Ge} \ , \\
&t_{1/2}^{(0\nu)} = (4.53-6.90)\times 10^{23}\,\textrm{yr}/
(\langle m_{\nu}\rangle[\textrm{eV}])^2 \textrm{ for}
\ \,^{82}\textrm{Se} \ .
\end{split} 
\end{equation}

In summary, we have calculated the nuclear matrix elements for the
$0\nu\beta\beta$ decays of $^{76}$Ge and $^{82}$Se by using the
proton-neutron quasiparticle random-phase approximation with a
realistic two-body interaction and a realistic single-particle space.
The numerical calculations were done by including the
higher-order terms of the nucleonic weak currents, the nucleon's
finite-size corrections and the nucleon-nucleon short-range
correlation effects. The short-range correlations have been calculated
by using the unitary correlation operator formalism that is superior
to the traditionally adopted rudimentary Jastrow procedure. The UCOM
reduces the bare values of the computed matrix elements less than the
Jastrow procedure, leading to larger matrix elements than the ones
quoted in the recent literature. This reduces the predicted
theoretical $0\nu\beta\beta$ half-lives of $^{76}$Ge and $^{82}$Se by
a significant amount and thus directly influences the neutrino-mass
sensitivities of the running and future double beta experiments.

This work has been partially supported by the 
Academy of Finland under the Finnish Centre of Excellence Programme 
2006-2011 (Nuclear and Accelerator Based Programme at JYFL). We thank
also the EU ILIAS project under the contract RII3-CT-2004-506222.


\begin{thebibliography}{99}
\bibitem{ELL02} S.R. Elliott and P. Vogel,
  Annu. Rev. Nucl. Part. Sci. {\bf 52}, 115 (2002). 
\bibitem{ELL04} S.R. Elliott and J. Engel, J. Phys. G {\bf 30}, R183
  (2004). 
\bibitem{BIL04} S.M. Bilenky and S.T. Petcov, hep-ph/0405237.
\bibitem{GEH07} V.M. Gehman and S.R. Elliott, hep-ph/0701099.
\bibitem{Oscillations} S. Fukuda {\it et al.}, Phys.
  Rev. Lett. {\bf 86}, 5651 (2001) ; Q.R. Ahmad {\it et al.}, Phys.
  Rev. Lett. {\bf 89}, 011301 (2002) ; H. Hagiwara {\it et al.}, Phys.
  Rev. D {\bf 66}, 010001 (2002) ; K. Eguchi {\it et al.}, Phys.
  Rev. Lett. {\bf 90}, 021802 (2003) ; M.H. Ahn {\it et al.}, Phys.
  Rev. Lett. {\bf 90}, 041801 (2003) ; S.N Ahmed {\it et al.}, Phys.
  Rev. Lett. {\bf 92}, 181301 (2004).
\bibitem{KLA01} H.V. Klapdor-Kleingrothaus {\it et al.},
  Mod. Phys. Lett. A {\bf 16}, 2409 (2001).
\bibitem{KLA04} H.V. Klapdor-Kleingrothaus {\it et al.},
  Phys. Lett. B {\bf 586}, 198 (2004).
\bibitem{ARN05a} R. Arnold {\it et al.}, Phys.
  Rev. Lett. {\bf 95}, 182302 (2005).
\bibitem{ARN05b} C. Arnaboldi {\it et al.}, Phys.
  Rev. Lett. {\bf 95}, 142501 (2005).
\bibitem{BAR04} A.S. Barabash, Phys. Atom. Nucl. {\bf 67}, 458 (2004).
\bibitem{DOI85} M. Doi, T. Kotani, and E. Takasugi, Prog.
  Theor. Phys. Suppl. {\bf 83}, 1 (1985).
\bibitem{PAS02} S. Pascoli, S.T. Petcov, and W. Rodejohann,
  Phys. Lett. B {\bf 549}, 2002 (177).
\bibitem{TOM91} T. Tomoda, Rep. Prog. Phys. {\bf 54}, 53 (1991).
\bibitem{SUH98} J. Suhonen and O. Civitarese, Phys. Rep. {\bf 300}, 123 (1998).
\bibitem{FAE98} A. Faessler and F. \v Simkovic, J. Phys. G {\bf 24},
  2139 (1998). 
\bibitem{CAU99} E. Caurier, F. Nowacki, A. Poves, and J. Retamosa,
  Nucl. Phys. A {\bf 654}, 973c (1999).
\bibitem{CAU05} E. Caurier {\it et al.}, Rev. Mod. Phys. 
  {\bf 77}, 427 (2005).
\bibitem{SUH07} J. Suhonen, {\em From Nucleons to Nucleus: Concepts
  of Microscopic Nuclear Theory} (Springer Verlag, Berlin, 2007).
\bibitem{TOI95} J. Toivanen and J. Suhonen, Phys.
  Rev. Lett. {\bf 75}, 410 (1995).
\bibitem{TOI97} J. Toivanen and J. Suhonen, Phys.
  Rev. C {\bf 55}, 2314 (1997).
\bibitem{ROD06} V.A. Rodin, A. Faessler, F. \v Simkovic, and P. Vogel,
  Nucl. Phys. A {\bf 766}, 107 (2006).
\bibitem{RQRPA} J. Engel {\it et al.}, Phys. Rev. C {\bf 55}, 1781
  (1997) ; A. Mariano and J.G. Hirsch, Phys. Rev. C {\bf 57}, 3015
  (1998) ; O. Civitarese and M. Reboiro, Phys. Rev. C {\bf 57}, 3062
  (1998). 
\bibitem{VOG86} P. Vogel and M.R. Zirnbauer, Phys. Rev. Lett. {\bf
    57}, 3148 (1986). 
\bibitem{CIV87} O. Civitarese, A. Faessler, and T. Tomoda,
  Phys. Lett. B {\bf 194}, 11 (1987).
\bibitem{SUH05} J. Suhonen, Phys. Lett. B {\bf 607}, 87 (2005).
\bibitem{CIV05} O. Civitarese and J. Suhonen, Phys. Lett. B {\bf 626},
  80 (2005), {\it ibid} Nucl. Phys. A {\bf 761}, 313 (2005).
\bibitem{HAX84} W.C. Haxton and G.J. Stephenson Jr.,
  Prog. Part. Nucl. Phys. {\bf 12}, 409 (1984).
\bibitem{ENG88} J. Engel, P. Vogel, and M.R. Zirnbauer, 
  Phys. Rev. C {\bf 37}, 731 (1988).
\bibitem{MIL76} G.A. Miller and J.E. Spencer, Ann. Phys. {\bf 100},
  562 (1976).
\bibitem{r} $r$ is the relative distance $r=|{\bf r}_{1}-{\bf r}_{2}|$ 
  between the nucleons 1 and 2.
\bibitem{rudi} This particular variant of the Jastrow function is
  quite rudimentary. Much more sophisticated, variational Jastrow 
  functions are used, e.g., in ab-initio calculations of
  the properties of very light nuclei.
\bibitem{FEL98} H. Feldmeier, T. Neff, R. Roth, and J. Schnack,
  Nucl. Phys. A {\bf 632}, 61 (1998).
\bibitem{KOR07} M. Kortelainen, O. Civitarese, J. Suhonen, and
  J. Toivanen, nucl-th/0701052, and to appear in Phys. Lett. B (2007).
\bibitem{ROT05} R. Roth {\it et al.}, Phys. Rev. C {\bf 72}, 034002
  (2005). 
\bibitem{ROT06} R. Roth {\it et al.}, Phys. Rev. C {\bf 73}, 044312
  (2006). 
\bibitem{SIM99} F. \v Simkovic, G. Pantis, J.D. Vergados, and
  A. Faessler, Phys. Rev. C {\bf 60}, 055502 (1999).
\bibitem{SUH88} J. Suhonen, T. Taigel, and A. Faessler, Nucl. Phys. A 
  {\bf 486}, 91 (1988).
\bibitem{SUH93} J. Suhonen, Nucl. Phys. A {\bf 563}, 205
  (1993), {\it ibid} {\bf 700}, 649 (2002).
\bibitem{BAR06} A.S. Barabash, Czech. J. Phys. {\bf 56}, 437 (2006). 
\bibitem{posint} Only positive values of the matrix elements were
  chosen by the beta-decay arguments of \cite{ROD06}.
\bibitem{RET95} J. Retamosa, E. Caurier, and F. Nowacki, Phys. 
  Rev. C {\bf 51}, 371 (1995). 
\bibitem{SUH90} J. Suhonen, S.B. Khadkikar, and A. Faessler,
  Phys. Lett. B {\bf 237}, 8 (1990), {\it ibid} Nucl. Phys. A 
  {\bf 529}, 727 (1991), {\it ibid} Nucl. Phys. A {\bf 535}, 509 (1991).
\bibitem{ALV04} R. \'Alvarez-Rodriguez {\it et al.}, Phys. Rev. C {\bf
    70}, 064309 (2004).
\bibitem{FRE07} S.J. Freeman {\it et al.}, nucl-ex/0701003.
\end{thebibliography}
\end{document}